\documentclass[12pt]{iopart}

\usepackage{graphicx}
\usepackage{dcolumn}
\usepackage{bm}
\usepackage{hyperref}
\usepackage{algpseudocode}
\usepackage{algorithm}
\usepackage{amsthm}
\usepackage{amssymb}
\usepackage{cleveref}
\usepackage{color}
\usepackage{cite}
\newcommand{\estimated}{\ensuremath{\mathbf{\hat{s}}}}
\newcommand{\measured}{\ensuremath{\mathbf{\tilde{s}}}}

\def\be{\begin{equation}}
\def\ee{\end{equation}}

\def\bc{\begin{center}}
\def\ec{\end{center}}
\def\bea{\begin{eqnarray}}
\def\eea{\end{eqnarray}}
\begin{document}

\title[Dirac signal processing of  higher-order topological signals]{Dirac signal processing of  higher-order topological signals}

\author{Lucille Calmon$^1$, Michael T. Schaub$^2$ and Ginestra Bianconi$^{1,3}$}
\address{$^1$School of Mathematical Sciences, Queen Mary University of London, London, E1 4NS, United Kingdom\\
$^2$Department of Computer Science,  RWTH Aachen University, Ahornstrasse 55, 52074 Aachen, Germany\\
$^3$The Alan Turing Institute, 96 Euston Rd, London NW1 2DB, United Kingdom}
\ead{ginestra.bianconi@gmail.com}
\vspace{10pt}
\begin{indented}
\item[]
\end{indented}

\begin{abstract}
Higher-order networks can sustain topological signals which are variables associated not only to the nodes, but also to the links, to the triangles and in general to the higher dimensional simplices of simplicial complexes. These topological signals can describe a large variety of real systems including currents in the ocean, synaptic currents between neurons and biological transportation networks.
In real scenarios topological signal data might be noisy and an  important task is to process these signals {by} improving their signal to noise ratio. So far topological  signals are typically processed independently of each other. For instance, node signals are processed independently of link signals, and algorithms that can enforce a consistent processing of topological signals across different dimensions are largely lacking.
Here we propose Dirac signal processing,  an adaptive, unsupervised signal processing algorithm that learns to jointly filter topological signals supported on nodes, links and  triangles of simplicial complexes in a consistent way. 
The proposed Dirac signal processing algorithm is formulated in terms of the discrete Dirac operator which can be interpreted as ``square root" of a higher-order Hodge Laplacian. We discuss in detail the properties of the Dirac operator including its spectrum and the chirality of its eigenvectors and we adopt this operator to formulate Dirac signal processing  that can filter noisy signals defined on nodes, links and triangles of simplicial complexes.
We test our algorithms on noisy synthetic data and noisy data of drifters in the ocean  and find that the algorithm can learn to efficiently reconstruct the true signals outperforming algorithms based exclusively on the Hodge Laplacian.
\end{abstract}

%
%
%
%
%

\section{Introduction}

Recently, higher-order network models~\cite{bianconi2021higher,battiston2020networks,battiston2021physics,bick2021higher,torres2021and} that capture many-body interactions within complex systems have gained increasing attention. 
Indeed, a vast number of systems ranging from the brain~\cite{giusti2016two,reimann2017cliques} to social and collaboration networks include interactions between two or more nodes that can be encoded in (simplicial) complexes and hypergraphs.
Higher-order networks and simplicial complexes not  {only reveal the topology of data \cite{giusti2016two,otter2017roadmap,xia2014persistent,petri2014homological} but have} also transformed  our understanding of the interplay between structure and dynamics~\cite{bianconi2021higher,battiston2021physics,majhi2022dynamics} in networked systems. 
In fact these higher order networks can support  topological signals {\cite{barbarossa2020topological,millan2020explosive,torres2020simplicial,Schaub2018b,schaub2020random,battiloro2023topological2}}, i.e.  variables not only associated to the nodes of a network but associated also to the links, to the (filled) triangles and in general to the higher dimensional simplices of simplicial complexes. 
Examples of topological signals include the number of citations that a team of collaborators have acquired in their co-authored papers, and link signals associated to the link of a network or of a simplicial complex.  {The latter  include{,} for instance}{,} synaptic signals and link-signals defined on links connecting  brain regions~\cite{faskowitz2022edges,santoro2022unveiling}.
Interestingly{,} vector fields, such as currents  at different locations in the ocean~\cite{schaub2020random} or speed of wind at a given altitude, can {also} be projected into a spatial tessellation and be treated as topological signals associated to the links of the tessellation.
These topological signals reveal important aspects of the interplay between structure and dynamics on networked structures, which can be fully accounted for only by using the mathematical tools of discrete topology \cite{grady2010discrete} and discrete exterior calculus  \cite{desbrun2005discrete}.
On one side it has been recently shown that topological signals undergo collective phenomena that display a rich interplay between topology and dynamics \cite{millan2020explosive,torres2020simplicial,ghorbanchian2021higher,calmon2022dirac,giambagli2022diffusion,calmon2022local,ziegler2022balanced,arnaudon2021connecting,ghorbanchian2022hyper}.
On the other side topological signal data can be processed by machine learning and signal processing algorithms \cite{Schaub2018b,barbarossa2020topological, schaub2020random,saito2022multiscale}, including neural networks \cite{bodnar2021weisfeiler,goh2022simplicial,giusti2022simplicial} which make use of  algebraic topology. 

In the last years signal processing on non-Euclidean domains such as graphs has received large attention~\cite{bronstein2017geometric}.
Using variants of the Hodge Laplacian operators~\cite{grady2010discrete,lim2020hodge}, these ideas have been extended to signal processing on simplicial complexes and other topological spaces{~\cite{barbarossa2020topological,Schaub2018b,Schaub2021,Schaub2021a,Yang2021,sardellitti2021topological,roddenberry2022signal,krishnagopal2021spectral,battiloro2023topological,battiloro2023topological2}} and found their way into extensions of graph neural networks{~\cite{roddenberry2019hodgenet,roddenberry2021principled,bodnar2021weisfeiler,ebli2020simplicial,hajij2020cell,saito2022multiscale,battiloro2023latent}}. 
Further{,} current approaches build both on the Hodge Laplacians or the Magnetic Laplacian~\cite{gong2021directed,zhang2021magnet,bottcher2022complex} to effectively extract information from higher-order or directed network data. 
However, despite great progress achieved so far, most current signal processing algorithms focus only on the processing of topological signals of a given dimension, e.g., flows on links.
Algorithms to jointly filter topological signals of different dimensions, e.g., to enforce certain consistency relationships between node and link signals, are currently mostly lacking.

In this work we fill this gap by proposing  {Dirac signal processing}, an algorithm which learns to jointly filter topological signals of different dimensions, coupled via appropriate boundary maps as encapsulated in the Dirac operator \cite{dirac,bianconi2022dirac}.
The discrete Dirac operator~\cite{dirac,bianconi2022dirac,lloyd2016quantum,ameneyro2022quantum,post2009first}, is defined as the sum between the exterior derivative and its dual and can intuitively be understood as a square root of the (Hodge) Laplacian operator. This operator has recently been shown to be key to treat dynamics of topological signals of different dimensions on networks, simplicial and cell complexes~\cite{dirac,giambagli2022diffusion,calmon2022dirac,calmon2022local}, 
To illustrate the utility of our approach we consider a signal smoothing task, in which a regularization based on the Dirac operator is used, to filter out noise in a topological signal in an adaptive and unsupervised fashion.
In particular, this improves upon an algorithm based on the normalized Dirac operator that we presented in a preliminary work~\cite{calmon2022higher}. {As detailed later on, the here proposed Dirac signal processing algorithm has the main advantage of adaptively learning its main parameter, as opposed to the preliminary method from~\cite{calmon2022higher}. This adaptability enables the algorithm to maintain a good performance over a wider range of signals than Ref.~\cite{calmon2022higher}. To show this, }
we test our  {Dirac signal processing} algorithm on both synthetic and real topological signals defined on nodes, links and (filled) triangles of networks and simplicial complexes. We find a very good performance of the algorithm under a wide set of conditions which typically outperform algorithms based exclusively on the Hodge Laplacian.

The paper is structured as follows. 
In~\Cref{sec:dirac_def} we provide the background on the Dirac operator and its spectral properties; in~\Cref{sec:smoothing} we formulate  {Dirac signal processing}; in~\Cref{sec:num_experiments} we discuss the result obtained by applying the proposed algorithm to synthetic and real data. 
Finally, in~\Cref{sec:conclusions} we provide some discussion and concluding remarks.
{Two Appendices containing details on the convergence  of the algorithm and on  its complexity follow.}

\section{The Dirac operator on simplicial complexes and its spectral properties}\label{sec:dirac_def}

\subsection{Simplicial complexes and topological spinors}

Simplicial complexes are higher-order networks \cite{bianconi2021higher} built from simplices. A $n$-dimensional simplex is a set of $n+1$ nodes{;} therefore{,} a $0$-simplex is a node, a $1$-simplex is a link, a $2$-simplex is a triangle{, }etc.
Note that for simplifying the terminology here and in the following we will always refer to filled triangles simply as triangles.
Here we consider simplicial complexes $\mathcal{K}$ of dimension $d=2$ formed exclusively by nodes, links and triangles which reduce to simplicial complexes of dimension $d=1$ (networks) in absence of triangles.
We indicate with  $N_{[n]}$ the number of simplices of dimension $n$, i.e. the considered simplicial complexes have $N_{[0]}$ nodes, $N_{[1]}$ links and $N_{[2]}$ triangles. 
In algebraic topology \cite{hatcher2005algebraic,bianconi2021higher}, the $n$-simplices are associated to an orientation, typically induced by the node labels.
In our signal processing setting, each simplex of the simplicial complex $\mathcal{K}$ is associated with a topological signal, which can describe either a node, a link or a triangle variable. 
Example of link signals are biological fluxes, or even electric currents.  In a scientific collaboration networks topological signals can encode{,} for instance{,} the number of citations that each single author, each pair of co-authors and each triple of co-authors has received from their joint works.


All the topological signals defined on a given simplicial complex are captured by a 
{\em topological spinor} $\mathbf s\in {C^0 \oplus C^1\oplus C^2}$ which is the direct sum of signal defined on nodes, links and triangles, i.e. {cochains in $C^n$ with $n=0,1,2$ respectively,} or equivalently,
\bea
\mathbf s=\left(\begin{array}{c}{\mathbf s}_0\\
{\mathbf s}_1\\{\mathbf s}_2\end{array}\right),
\eea
where ${\mathbf s}_n\in \mathbb{R}^{N_{[n]}}$.
Hence the topological spinor encodes all signals supported on the $d=2$ dimensional simplicial complex $\mathcal K$.

\subsection{The Dirac operator}
The  Dirac operator ${\bf D}$ is a linear operator that acts on the topological spinor~\cite{dirac,baccini,topsynchr,giambagli2022diffusion,bianconi2022dirac}. 
In the canonical basis of topological spinors of a $d=2$ dimensional simplicial complex, the  Dirac operator ${\bf D} $ has the block structure  
\begin{equation}
{\bf D}=\left(\begin{array}{ccc}0& {\bf B}_{[1]}& 0\\
{\bf B}_{[1]}^{\top}& 0 &{\bf B}_{[2]}\\
0& {\bf B}_{[2]}^{\top} &0\end{array}\right)
\end{equation}
where ${\bf B}_{[n]}$ is the $N_{[n-1]}\times N_{[n]}$  boundary matrix of order $n\ge 1$ and $ {\mathbf B}_{[n]}^{\top}$ the coboundary matrix (for the definition of boundary matrix see for instance \cite{bianconi2021higher}).

 We notice that the  Dirac operator can be considered the  ``square root'' of the super-Laplacian operator $\mathcal{L}$, i.e.
\bea
{{\bf D}}^2=\mathcal{L}=\left(\begin{array}{ccc} {\bf L}_{[0]}& 0&0\\
 0 &{\bf L}_{[1]}&0\\
0& 0& {\bf L}_{[2]}\end{array}\right),
\label{HodgeL}
\eea
where ${\bf L}_{[n]}$ indicates the Hodge Laplacian of dimension $n$, where
\bea
{\bf L}_{[0]}={\bf B}_{[1]}{\bf B}_{[1]}^{\top},
\eea
and for $n\geq 1$
\bea
{\bf L}_{[n]}={\bf B}_{[n]}^{\top}{\bf B}_{[n]}+{\bf B}_{[n+1]}{\bf B}_{[n+1]}^{\top}.
\eea
Note that for simplicial complexes of dimension $d=2$ we put ${\bf B}_{[3]}={\bf 0}$ while for simplicial complexes of dimension $d=1$ (i.e. networks) we put also ${\bf B}_{[2]}={\bf 0}$.

The Hodge Laplacian ${\bf L}_{[n]}$ describes diffusion-like dynamics~\cite{torres2020simplicial,ziegler2022balanced} occurring among $n$-dimensional simplices and can be decomposed into two terms ${\bf L}_{[n]}={\bf L}_{[n]}^{up}+{\bf L}_{[n]}^{down}$ with 
 ${\bf L}_{[n]}^{up}={\bf B}_{[n+1]}{\bf B}_{[n+1]}^{\top}$ and ${\bf L}_{[n]}^{down}= {\bf B}_{[n]}^{\top}{\bf B}_{[n]}$  describing diffusion through $(n+1)$-dimensional simplices and $(n-1)$-dimensional simplices respectively.
 
 An important topological property is that the boundary of a boundary is null, hence ${\bf B}_{[n]}{\bf B}_{[n+1]}={\bf 0}$ and  ${\bf B}_{[n+1]}^{\top}{\bf B}_{[n]}^{\top}={\bf 0}$ with $n\geq 1$.

 This property has the direct consequence that \bea
 {\bf L}_{[n]}^{down}{\bf L}_{[n]}^{up}={\bf L}_{[n]}^{up}{\bf L}_{[n]}^{down}={\bf 0},
 \eea
 which implies
 \bea
 \mbox{im}({\bf L}_{[n]}^{down})\subseteq \mbox{ker}({\bf L}_{[n]}^{up}),\quad
  \mbox{im}({\bf L}_{[n]}^{up})\subseteq \mbox{ker}({\bf L}_{[n]}^{down}).\quad
 \eea
 
 Therefore any signal supported on the $n$-dimensional simplices of a simplicial complex can be uniquely decomposed into the sum of three terms: one in the kernel of ${\bf L}_{[n]}$, one in the image of ${\bf L}_{[n]}^{up}$ and one in the image of ${\bf L}_{[n]}^{down}$, 
 \bea
 \mathbb{R}^{N_{[n]}}=\mbox{ker}({\bf L}_{[n]})\oplus \mbox{im}({\bf L}_{[n]}^{up})\oplus \mbox{im}({\bf L}_{[n]}^{down}).
 \eea
 This decomposition is called {\em Hodge decomposition} and is a central result in algebraic topology.
 For signals defined on links, this implies a unique decomposition into an harmonic, a solenoidal and an irrotational component \cite{bianconi2021higher,hatcher2005algebraic}.

 \subsection{Dirac decomposition}
 Similar to the Hodge decomposition a signal decomposition of the topological spinors based on the Dirac operator can be derived.
 For this, note first that the Dirac operator can be written as 
\bea
{\bf D}={\bf D}_{[1]}+{\bf D}_{[2]}
\eea
where ${\bf D}_{[n]}$ are defined as 
\bea
{\bf D}_{[1]}=\left(\begin{array}{ccc}0& {\bf B}_{[1]}& 0\\
{\bf B}_{[1]}^{\top}& 0 &0\\
0& 0 &0\end{array}\right), \quad 
        {\bf D}_{[2]}=\left(\begin{array}{ccc}0& 0& 0\\
0& 0 &{\bf B}_{[2]}\\
0& {\bf B}_{[2]}^{\top} &0\end{array}\right).\quad
\eea

This implies the following properties.
First, the squares of ${\bf D}_{[n]}$ are given by  
\bea
{\bf D}_{[1]}^2=\mathcal{L}_{[1]}=\left(\begin{array}{ccc} {\bf L}_{[0]}& {\bf 0} &{\bf 0}\\
{\bf 0}&{\bf L}_{[1]}^{down}& {\bf 0}\\
{\bf 0}& {\bf 0} &{\bf 0}\end{array}\right), \nonumber \\
{\bf D}_{[2]}^2=\mathcal{L}_{[2]}=\left(\begin{array}{ccc}{\bf 0}& {\bf 0}& {\bf 0}\\
{\bf 0}& {\bf L}_{[1]}^{up} &{\bf 0}\\
{\bf 0}& {\bf 0}&{\bf L}_{[2]}^{down} \end{array}\right).
\label{D1L}
\eea

Secondly we note that ${\bf D}_{[1]}$ and ${\bf D}_{[2]}$ commute and satisfy 
\bea
{\bf D}_{[1]}{\bf D}_{[2]}= {\bf D}_{[2]}{\bf D}_{[1]}={\bf 0}
\eea
which  implies
\bea
\mbox{im}({\bf D}_{[2]})\subseteq\mbox{ker}({\bf D}_1),\quad\mbox{im}({\bf D}_{[1]})\subseteq\mbox{ker}({\bf D}_{[2]}).
\eea
Stated differently, we have what we call a {\em Dirac decomposition}:  
\bea
\mathcal{C}=\mbox{ker}({\bf D})\oplus\mbox{im}({\bf D}_{[1]})\oplus\mbox{im}({\bf D}_{[2]})
\label{Dirac_dec}.
\eea

Note that the kernel of the Dirac operator,  $\mbox{ker}({\bf D})$ is given by 
\bea
\mbox{ker}({\bf D})=\mbox{ker}({\bf L}_{[0]})\oplus \mbox{ker}({\bf L}_{[1]})\oplus \mbox{ker}({\bf L}_{[2]})
\eea
and its dimension is therefore given by the sum of the Betti numbers $\beta_n$ of the simplicial complex:
\bea
\mbox{dim ker}({\bf D})=\beta_0+\beta_1+\beta_2.
\eea
From the Dirac decomposition it  follows that a signal $\bf s$ defined on nodes, links and triangles  can be decomposed uniquely into the sum of an harmonic signal $s_{\mbox{harm}}\in \mbox{ker}({\bf D})$ and two signals $\bf s_{[1]}$ in the image of ${\bf D}_{[1]}$ and ${\bf s_{[2]}}$ in the image of ${\bf D}_{[2]}$, respectively:
\begin{equation}
\label{eq:dirac_decomp}
\mathbf s=\mathbf s_{[1]}+\mathbf s_{[2]}+\mathbf s_{\mbox{harm}}.
\end{equation}
 {Note that $\bm s_{[1]}$ is a topological spinor of non-zero elements on both nodes and links while $\bm s_{[2]}$ is a topological spinor of non-zero elements on both links and triangles.}
The two signals ${\bf s}_{[n]}$ with $\mathbf s_{[n]}\in \mbox{im}({\bf D}_{[n]})$ can be written as
\bea
{\bf s}_{[n]}={\mathbf{D}}_{[n]} \mathbf w_{[n]}, \quad
\eea
where $\mathbf w_{[n]}$ can be found by minimizing 
\bea
\mathbf w_n=\mbox{argmin}_{\mathbf w_{[n]}}\|{\bf D}_{[n]}\mathbf w_{[n]}-\mathbf s\|_2,
\eea
which leads to 
\bea
{\mathbf w}_{[n]}={\mathbf D}_{[n]}^{\dag}{\mathbf s},
\eea
where ${\bf D}_{[n]}^{\dag}$ indicates the pseudoinverse of ${\bf D}_{[n]}$.
It follows that ${\mathbf s}_{[1]}$ and ${\mathbf s}_{[2]}$ can be expressed as 
\bea
{\mathbf s}_{[n]}={\mathbf D}_{[n]}{\mathbf D}_{[n]}^{\dag}{\mathbf s},
\label{projection}
\eea
with $n\in \{1,2\}$.

\subsection{Eigenvalues and eigenvectors of the Dirac operator}

Let us now exploit the fact that ${\bf D}_{[n]}^2=\mathcal{L}_{[n]}$, i.e. Eq. (\ref{D1L}) to relate the eigenvalues of the Dirac operators ${\bf D}_{[n]}$ to the eigenvalues of the Hodge Laplacians ${\bf L}_{[n-1]}^{up}$ and ${\bf L}_{[n]}^{down}$. Since ${\bf L}_{[n-1]}^{up}$ and ${\bf L}_{[n]}^{down}$ are isospectral, i.e. they have the same non-zero spectrum, from Eq. (\ref{D1L}) it follows the non-zero eigenvalues of ${\mathbf D}_{[n]}$ indicated as $\lambda$ satisfy 
\bea
\lambda=\pm\sqrt{\mu},
\eea
where $\mu$ are the non-zero eigenvalues of ${\bf L}_{[n-1]}^{up}$. Therefore for any positive eigenvalue $\lambda$ the operator ${\bf D}_{[n]}$ admits a negative eigenvalue $\lambda^{\prime}=-\lambda$.

The eigenvectors associated to non-zero eigenvalues of the operator ${\bf D}_{[n]}$ are related by chirality for both $n=1,2$.
To see this, let us introduce the matrices $\bm \gamma_{[n]}$ 
given by 
\bea
\bm\gamma_{[1]}=\left(\begin{array}{ccc}{\bf I}&{\bf 0}&{\bf 0}\\{\bf 0}& -{\bf I}& {\bf 0}\\
{\bf 0}&{\bf 0}&{\bf 0}
\end{array}\right),\quad
\bm\gamma_{[2]}=\left(\begin{array}{ccc}{\bf 0}&{\bf 0}&{\bf 0}\\{\bf 0}& {\bf I}& {\bf 0}\\
{\bf 0}&{\bf 0}&-{\bf I}
\end{array}\right).\nonumber \\
\eea
It is easy to check that the Dirac operator ${\bf D}_{[n]}$ anticommutes with $\bm \gamma_{[n]}$, i.e.
\bea
\{{\bf D}_{[n]},\bm\gamma_{[n]}\}={\bf D}_{[n]}\bm\gamma_{[n]}+\bm\gamma_{[n]}{\bf D}_{[n]}=\mathbf 0.
\eea

Hence if we denote by $\bm \phi^{+}_{n}$ an eigenvector of ${\bf D}_{[n]}$ with eigenvalue $\lambda>0$ we assume that
\bea
{\bf D}_{[n]}\bm\phi^{+}_n=\lambda\bm\phi_n^{+},
\eea 
it follows from direct computation that  $\bm\gamma_{[n]}\bm\phi^{+}_n$ is an eigenvector of ${\bf D}_{[n]}$ with eigenvalue $\lambda'=-\lambda$:
\bea
{\bf D}_{[n]}(\bm\gamma_{[n]}\bm\phi_n^{+})=-\bm\gamma_{[n]}{\bf D}_{[n]}\bm\phi_n^{+}=-\lambda (\bm\gamma_{[n]}\bm\phi_n^{+}).
\eea
This proves that for any  eigenvector $\bm\phi^{+}_n$ of ${\bf D}_{[n]}$, corresponding to a positive eigenvalue $\lambda$  there is an eigenvector $\bm\phi^{-}_n=\bm\gamma_{[n]}\bm\phi^{+}_n$ corresponding to the negative eigenvalue $\lambda'=-\lambda$.

The Dirac decomposition given by Eq.(\ref{Dirac_dec}) implies that non-zero eigenvalues of the Dirac operator ${\bf D}$ are either non-zero eigenvalues of ${\bf D}_{[1]}$ or non-zero eigenvalues of ${\bf D}_{[2]}$. Moreover the non-zero eigenvectors of ${\bf D}$ are either non-zero eigenvectors of ${\bf D}_{[1]}$  or non-zero eigenvectors of ${\bf D}_{[2]}$.
Therefore the matrix $\bm \Phi$ of the eigenvectors of ${\bf D}$ can be written as 
\bea
    \bm \Phi = 
    \left(\begin{array}{ccc}\bm \Phi_{[1]}& \bm \Phi_{[2]} &\bm \Phi_{\mbox{harm}} \\
    \end{array}\right).
\eea
where ${\bm \Phi}_{[n]}$ is the matrix of the eigenvectors ${\bm\phi}_n\in \mbox{im}({\bf D}_{[n]})$ with $n\in \{1,2\}$ and $\bm\Phi_{\mbox{harm}}$ is the matrix of eigenvectors forming a basis for $\mbox{ker}({\bf D})$. 
The generic (non-normalized) eigenvectors $\bm \phi_n^{+}$ and $\bm \phi_n^{-}$ associated to positive and negative eigenvalues of the Dirac operator $\mathbf D_{[n]}$ respectively have opposite chirality and can be written as 
\bea
\bm \phi_1^+=\left(\begin{array}{c} {\bf u}_1\\  {\bf v}_1\\{\bf 0}\end{array}\right),\quad \bm \phi_1^-=\left(\begin{array}{c} {\bf u}_1\\  -{\bf v}_1\\{\bf 0}\end{array}\right)
\nonumber \\
\bm \phi_2^+=\left(\begin{array}{c}{\bf 0}\\{\bf u}_2 \\   {\bf v}_2 \end{array}\right), \quad \bm \phi_2^-=\left(\begin{array}{c}{\bf 0}\\{\bf u}_2 \\   -{\bf v}_2 \end{array}\right),
\label{eq:phipm}
\eea
where ${\bf u}_n, {\bf v}_n$ indicate the left and right singular vectors of ${\mathbf B}_{[n]}$.
Let us indicate  with  ${\mathbf U}_{[n]},{\mathbf V}_{[n]}$, the matrices   formed by left (${\mathbf U}_{[n]}$) and right (${\mathbf V}_{[n]}$) singular vectors of ${\mathbf B}_{[n]}$ .

We can then assemble a matrix of (non-normalized) eigenvectors  ${\bm \Phi_{[n]}}$ with $n\in \{1,2\}$ in the form 
\bea
    &\bm \Phi_{[1]} = 
    \left(\begin{array}{cc}{\mathbf U}_{[1]} & {\mathbf U}_{[1]}  \\
    {\mathbf V}_{[1]}  & -{\mathbf V}_{[1]} \\
    \mathbf 0&\mathbf 0
    \end{array}\right),
      \bm \Phi_{[2]} = 
    \left(\begin{array}{cc}\mathbf 0&\mathbf 0\\{\mathbf U}_{[2]} &  {\mathbf U}_{[2]}  \\
    {\mathbf V}_{[2]}  & -{\mathbf V}_{[2]} \\
    \end{array}\right).
\eea
Importantly, there is always a basis in which the harmonic eigenvectors are localized only on $0, 1$ and $2$ dimensional simplices, i.e.we have:
\bea
\bm \Phi_{\mbox{harm}}=\left(\begin{array}{ccc} {\bf U}_{\mbox{harm}}&\mathbf 0 & \mathbf 0 \\
\mathbf 0 & {\bf V}_{\mbox{harm}}&\mathbf 0\\
\mathbf 0 &\mathbf 0 &  {\bf Z}_{\mbox{harm}}
    \end{array}\right).
\eea
We remark that it is further possible to define a (weighted) normalized Dirac operators~\cite{baccini,calmon2022higher}. 
In this paper we focus exclusively on the unnormalized Dirac operator. 
However, all algorithms we derive in the following can be also be formulated in terms of the normalized Dirac operator.

\section{Dirac signal processing}\label{sec:smoothing}

\subsection{Problem Setup and  {Dirac signal processing} algorithm}
To illustrate how the Dirac operator can be used for joint signal processing of topological signals, we consider a filtering problem.
Specifically, we are interested in a scenario in which we observe a noisy signal:
\begin{equation}
    \measured = \mathbf s + \bm \epsilon 
\end{equation}
where $\mathbf s$ is the true signal vector we wish to reconstruct and $\bm \epsilon$ is noise.
For simplicity, we consider the case in which the true signal has norm one $\|\mathbf s\|_2=1$.

Most works in the literature so far (see, e.g.,~\cite{Schaub2021,Schaub2021a} for an overview) have considered this problem setup from a smoothing perspective, using the assumption that $\mathbf{s}$ is approximately harmonic.
In this case, one can use an optimization formulation of the following form:
\begin{equation}
\min_{\estimated} \left [ \left(\|\estimated -\measured \|_2\right)^2 + \tau \estimated^\top\mathcal L\estimated\right ],
\end{equation}
{ with $\cal{L}$ defined in Eq. (\ref{HodgeL}) and whose solutions is given by}
\begin{equation}
\estimated=[{\mathbf I}+\tau \mathcal{L}]^{-1}\measured,    
\end{equation}
to yield an estimate $\estimated$ which acts as a ``low pass'' filter and attenuates signal components associated to eigenvectors with high eigenvalues (high frequencies) of $\mathcal{L}$, while attenuating signal components associated to eigenvectors with small eigenvalues (low frequences) less, and leaving harmonic components completely unaltered.
As the eigenvalues of the above filter are of the form $1/(1+\tau \lambda_i)$, the parameter $\tau$ regulates the steepness of this attenuation: the larger $\tau$ the more the higher-frequencies (eigenvalues) are attenuated and filtered out.

In contrast, in the following we are interested in a more general problem setup, in which we know that the true signal is effectively localized in a subspace spanned by eigenvectors of the Dirac (respectively, super-Laplacian) operator, but we \emph{do not know} which subspace this is. 
In other words, we do not know \emph{a priori} to which eigenvalues (frequencies) the relevant eigenvectors are associated, but we nevertheless want to establish an algorithm that can filter in an unsupervised way the relevant components.

In our scenario  we will  consider that our true signal is orthogonal to the harmonic space,  i.e. $\mathbf s_{\mbox{harm}}=\mathbf 0$.
{This might not always be the case, but it is a scenario in which the most commonly used low pass filters are less effective.}

We now observe that, since all the considered vectors ${\mathbf s}$, $\measured$ and $\bm\epsilon$ belong to  the space of topological spinors $\mathcal{C}$ defined in Sec. \ref{sec:dirac_def}, according to Dirac decomposition (see Eq. (\ref{eq:dirac_decomp})) there is a unique way to write them as
\bea
\measured&=&\measured_{[1]}+\measured_{[2]}+\measured_{\mbox{harm}}\nonumber\\
\mathbf s&=&\mathbf s_{[1]}+\mathbf s_{[2]}+\mathbf s_{\mbox{harm}},\nonumber\\
\bm\epsilon&=&\bm\epsilon_{[1]}+\bm\epsilon_{[2]}+\bm\epsilon_{\mbox{harm}}
\eea
where $\measured_{[n]}, {\mathbf s}_{[n]}$ and $\bm \epsilon_{[n]}$ can be calculated using Eq. (\ref{projection}). 
It follows that our task can be decomposed equivalently into the problem of reconstructing the two signal vectors $\mathbf s_{[1]}$  and $\mathbf s_{[2]}$ from 
\bea
\measured_{[1]}=\mathbf s_{[1]}+\bm\epsilon_{[1]},\nonumber \\
\measured_{[2]}=\mathbf s_{[2]}+\bm\epsilon_{[2]},
\eea
where we indicate the reconstructed signal as 
\bea
\estimated=\estimated_{[1]}+\estimated_{[2]}.
\eea
Here, $\estimated_{[1]}$ is a signal associated to $\mbox{im}(\mathbf{D}_{[1]})$ and supported on nodes and links only, and   $\estimated_{[2]}\in \mbox{im}({\bf D}_{[2]})$ is a signal solely supported on links and triangles.
Similar to the low-pass setting, to estimate $\mathbf s_{[n]}$ based on the observed signals $\measured_{[1]}$ and $\measured_{[2]}$ we now propose to solve an optimization problem of the following form
\begin{equation}
\min_{\estimated_{[n]}} \left [ \|\estimated_{[n]} -\measured_{[n]} \|_2^2 + \tau \estimated_{[n]}^\top \mathbf Q_{[n]} \estimated_{[n]}\right ].
\label{eq:opt_problem}
\end{equation}
where now, however, $\mathbf{Q}_{[n]}$ acts as a quadratic regularizer that depends on a matrix polynomial $\mathbf Q_{[n]}$ of the Dirac operator of order $K$:
\begin{equation}
    {\mathbf Q}_{[n]} = \sum_{j=0}^K a_j{\mathbf D}^j_{[n]}.
\end{equation}
Here $a_i$ are coefficients, which have to be chosen such that ${\mathbf Q}_{[n]}$ is positive (semi-)definite to yield a well defined problem.
Note that for $a_2=1$ and $a_j=0$ for $j\neq 2$ this reduces to the block-diagonal standard Hodge-Laplacian kernel $\mathcal{L}$~\cite{barbarossa2020topological,Schaub2021a} (see Eq. (\ref{HodgeL})), and thus the topological signals of different dimension are filtered \emph{independently} in this setting.
\begin{figure*}[!htb]
\centering
\includegraphics[width=\columnwidth]{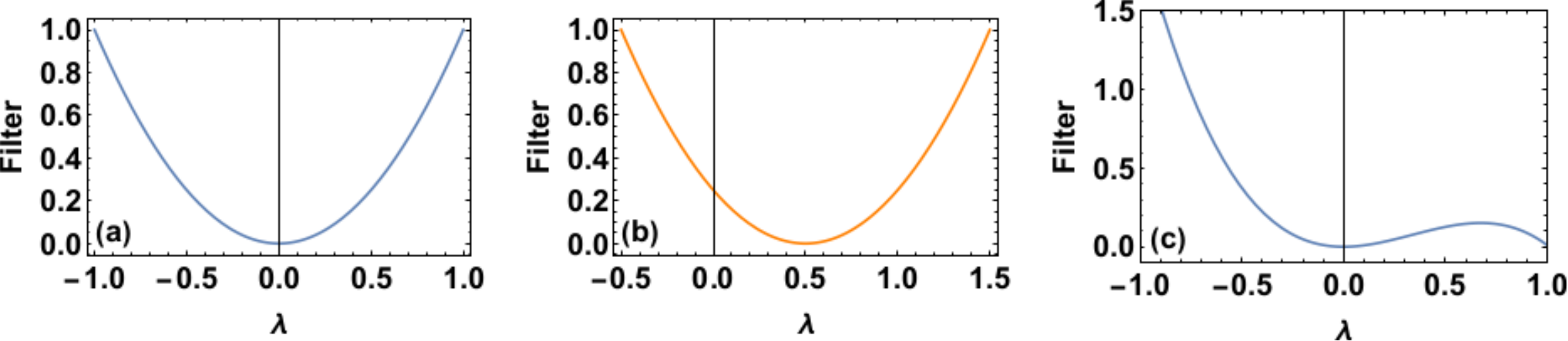}
\caption{Schematic representation on how the regularization term acts on the eigenmodes of the Dirac operator associated to the eigenvalue $\lambda$ for different choices of the regularization kernel ${\mathbf Q}_{[n]}$. Panel (a) represents the Hodge Laplacian kernel $Q_{[n]}=({\mathbf D}_{[n]})^2$ corresponding to $m_n=0$, panel (b) represents  the Dirac signal processing kernel $Q_{[n]}=({\mathbf D}_{[n]}-m_n{\bf I})^2$ adopted in this study here with $m_n=0.5$, while panel (c) represent the cubic kernel $Q_{[n]}=(\hat{{\mathbf D}}_{[n]})^2-z(\hat{{\mathbf D}}_{[n]})^3$ adopted in Ref. \cite{calmon2022higher}. {Note that in the derivation for this filter~\cite{calmon2022higher} it was assumed that one considers the normalized Dirac operator $\hat{{\mathbf D}}_{[n]}$ having eigenvalues $|\lambda|\leq 1$ (here represented for $z=0.99$).}}
\label{fig:Dirac_filters}
\end{figure*}
In the following we focus on a particular positive definite choice of $\mathbf Q_{[n]}$ suitable for our task of filtering out signal components with an adjustable frequency (associated eigenvalue):
 \bea
 \mathbf Q_{[n]}(m_n)=({\mathbf D}_{[n]}-m_{n}{\bf I})^2,
 \eea
 where  $m_{n}$ is an \emph{a priori} unknown constant that can be used to tune the filtering procedure.  
The solution of the optimization problem defined in Eq.(\ref{eq:opt_problem}) is then
\bea
\estimated_{[n]}&=&[{\mathbf I}+\tau  ( {\bf D}_{[n]} - m_n{\mathbf I})^2]^{-1}\measured_{[n]}.\label{eq:sol_tr}
\eea
which shows that signals aligned with eigenvectors corresponding to eigenvalues $\lambda\sim m_n$ will be attenuated least.
Observe that, indeed, if we set $m_1=m_2=0$, we recover a standard Hodge Laplacian kernel, which promotes harmonic signals (see panel (a) and (b)  of Fig. \ref{fig:Dirac_filters} for a schematic representation of the Hodge Laplacian and the Dirac regularization terms).

An essential question now is, of course, how we should set $m_{n}$ in the absence of a priori information about the true signal.
Here we utilize the following observation to develop our  {Dirac signal processing} algorithm.
If we had access to the true signals $\mathbf{s}_{[n]}$, and we knew that these were associated to predominantly to a specific eigenvalue $\lambda$, then we should set $m_n=\lambda$. 
Equivalently, we could use the information about $\mathbf{s}_{[n]}$ and $\mathbf{D}_{[n]}$ to compute $m_n$ as:
\bea
m_n=\frac{{\mathbf s}^{\top}_{[n]}{\bf D}_{[n]}{\mathbf s}_{[n]}}{\mathbf s^{\top}_{[n]}\mathbf s_{[n]}} = \lambda, 
\eea
which follows from simple computations (observe that we are effectively computing a Rayleigh quotient).
Crucially, while the true signal components are of course not available, for not too large signal-to-noise ratios, we may compute a proxy of the above statistic from the observed data.

The optimization problem  (\ref{eq:opt_problem}) can be then solved iteratively, by first solving the optimization problem finding $\estimated_{[n]}$ with the currently estimated estimated value $\hat{m}_n$, and then refining the estimate $\hat{m}_n$ using $\estimated_{[n]}$, with 
\bea
\estimated_{[n]}&=&[{\mathbf I}+\tau  ( {\bf D}_{[n]} - \hat{m}_n{\mathbf I})^2]^{-1}\measured_{[n]},\nonumber \\
\hat{m}_n&=&\frac{\estimated_{[n]}^{\top} \mathbf D_{[n]} \estimated_{[n]}}{\estimated_{[n]}^{\top}\estimated_{[n]}}.
\eea

This leads to the following unsupervised \emph{ {Dirac signal processing}} algorithm:
\begin{algorithmic}
\Require  Initial guess $\hat{m}_n^{(0)}$, Convergence threshold $\delta$, Learning rate $\eta$, Measured data $\measured_{[n]}$
\State $t \gets 0$
\State $\hat{m}_n(t=0) \gets \hat{m}_n^{(0)}$
\While{$|\hat{m}_n(t)-\hat{m}_n(t-1)|<\delta$}
	\State $t \gets t+1$
    \State $\estimated_{[n]} \gets [{\bf I}-\tau ({\bf D}_{[n]}-m_n{\bf I})^2]^{-1}\measured_{[n]}$ 
    \State $\hat{m}_n(t+1) \gets (1-\eta)\hat{m}_n(t)+\eta \frac{\estimated_{[n]}^{\top}{\bf D}_{[n]}\estimated_{[n]}}{\estimated_{[n]}^{\top}\estimated_{[n]}}$
\EndWhile
\end{algorithmic}
Here we indicate the iteration of the algorithm with $t$, and require as parameters a convergence threshold $\delta$, an initial guess for $m_n$, denoted by $\hat{m}_n^{(0)}$, and need to set a learning rate $0<\eta\leq 1$ for $\hat{m}_n$.

Note that the initial guess $\hat{m}_n^{(0)}$ may of course be computed by a Rayleigh coefficient using the observed data as well.
This strategy is in particular well suited if the signal-to-noise ratio is reasonably large; for a very low signal to noise ratio a good guess can be crucial to ensure effective convergence to a good approximation of the true $m_n$, however {(see Appendix A for details)}.

 An intuitive description of how the algorithm works is as follows.
 As discussed, the parameter $m_n(t)$ serves as an estimate of the  eigenvalue(s) of the dominant eigenvector contribution(s) that can be found in the signal. 
 This estimate $m_n(t)$ will then be used for a filtering round and thus will attenuate the frequencies around $m_n(t)$ the least.
 For a sufficiently close guess, this will result in an even better estimate of the dominant signal part, and thus will lead to a convergence of the algorithm, by {``}locking in{''} the desired frequency components automatically in a data driven way.
 We remark that in practice some deviation of the true signal from a single frequency is tolerable, as long as the set of relevant eigenvalues remains reasonably compact.

If the true signal $\mathbf s_{[n]}$ is known, the performance of the algorithm can be evaluated by monitoring the error 
\bea
\Delta  s_{n} = \|\estimated_{[n]} - \mathbf s_{[n]}\|_2,
\label{eq:er}
\eea
within the loop as a function of the iteration count $t$.

Note that the algorithm above is adaptive in that the filtering will automatically adjust according to the initial input provided.
This contrasts with the preliminary work we presented in~\cite{calmon2022higher}, in which we further adopted a different (fixed) regularizer and worked with the normalized Dirac operator instead.

The main difference of the present algorithm with respect to the one we proposed in Ref.\cite{calmon2022higher} are: (i)  In \cite{calmon2022higher} the signal processing algorithm uses the symmetrized version of the normalized Dirac operator defined in Ref. \cite{baccini} instead of the unnormalized Dirac operator used here; (ii) the regularization term used in Ref. \cite{calmon2022higher} uses $\bm Q=\hat{\bf D}^2-z\hat{\bf D}^3$; (iii) the algorithm proposed in \cite{calmon2022higher} does not learn any of its parameters.
Overall Dirac signal processing provides an improvement with respect to the algorithm proposed in \cite{calmon2022higher}.
The choice adopted here for the regularization term allows for a good performance of the Dirac signal processing on synthetic signals formed by arbitrary eigenvectors of the Dirac operators while the algorithm proposed in \cite{calmon2022higher} is best suited to treat true signals constituted by the eigenvector corresponding to the largest positive eigenvalue of the Dirac operator or by its chiral eigenvector. Indeed by tuning the value of $m$ the regularization term smooths the signal by selecting the eigenmodes around $m_n$ while the regularization term $\bm Q=\hat{\bf D}^2-z\hat{\bf D}^3$ has two minima that impede the reconstruction of arbitrary signals (see schematic representations of the filters in Fig. $\ref{fig:Dirac_filters}$).
Moreover  the proposed Dirac signal processing is an algorithm that learns to efficiently filter the noisy signal, which is a clear advantage in an unsupervised framework.
Finally we note that  Dirac signal processing  algorithm  can be easily generalized  by adopting  the  normalized Dirac operator instead of the unnormalized one, although in the cases analysed in this paper we have found no improvement of the performance of the algorithm.
On the contrary the algorithm proposed in \cite{calmon2022higher} can only be defined using the normalized Dirac operator.

\subsection{Numerical Experiments}
We conducted several numerical experiments on synthetic and real world data as described in the following.

\subsubsection{Synthetic signals}
For synthetic data on networks and simplicial complexes in which the true (or noisy) topological signals are not available, the signal $\mathbf s$ is taken to be the linear composition of two signals $\mathbf s_{[1]}$ and $\mathbf s_{[2]}$ with $\mathbf s_{[n]}$   aligned with a single  eigenvector of ${\mathbf D}_{[n]}$ corresponding to a non-degenerate eigenvalue (either $\bm \phi_n^{+}$ or $\bm\phi^{-}_n$ for any choice of $n\in \{1,2\}$).

In particular we have 
\bea
\mathbf s=\mathbf s_{[1]}+\mathbf s_{[2]}
\eea
with 
\bea
\mathbf s_{[1]}= \bm \phi_1^{\pm},\quad \mathbf s_{[2]}= \bm \phi_2^{\pm},
\label{eq:signals_eig}
\eea
where $\bm \phi_{n}^{\pm}$ are defined in Eq. (\ref{eq:phipm}).
Note that choosing both $\mathbf s_{[1]}$ and $\mathbf s_{[2]}$ proportional to eigenvectors associated to positive eigenvalues is a useful convention but the method works also with any other combination of eigenvectors associated to positive and negative eigenvalues.

Additionally, we can consider signals $\mathbf s_{[n]}$ built from linear combination of the eigenvectors of the Dirac operator with Gaussian coefficients,   i.e.
\bea
\mathbf s_{[n]}=\sum_{\lambda\neq  0} c_{\lambda}\bm\phi_n(\lambda)
\eea
where $\bm \phi_n(\lambda)$ indicates the eigenvector of ${\bf D}_{[n]}$ with eigenvalue $\lambda$ and  $c_{\lambda}=1/\mathcal{Z}\exp\left({-(\lambda-\bar{\lambda})^2/{2\hat{\sigma}}}\right)$ where $\mathcal{Z}$ is a normalization constant ensuring $\|\mathbf s_{[n]}\|_2=1$ and $\bar{\lambda}, \hat{\sigma}$ are two parameters determining $\mathbf s_{[n]}$.

\subsubsection{Signals generated from real-world data}

For several real-world datasets there are topological signals available only for a subset of dimensions of interest.
For instance, we might have network datasets formed by nodes and links where only node signals, or only link signals are available. 
Analogously, for a simplicial complex of dimension two, formed by nodes, links and triangles, we may have only access to signals supported on the links.
In these cases we can use the available real-world data to generate synthetic datasets as follows.
To generate the signals for the dimensions in which we have no measurements, we simply apply the Dirac operator to the observed signal, thus generating signals in all adjacent dimensions.
More precisely, let $\bm \sigma$ indicate the topological spinor defined on nodes, links and triangles, that has non-zero elements only in one observed dimension.
We then set our signal to:
\bea
\mathbf s_{[n]} =c_n(\bm \sigma+{\bf D}_{[n]} \bm \sigma),
\label{eq:dr}
\eea
where $c_n$ is a normalization constant that we set to enforce the condition $\|\mathbf s_{[n]}\|_2=1$. 
Hence, if $\bm \sigma$ is defined only on nodes $\mathbf s =\mathbf s_{[1]}+\mathbf s_{[2]}$ will be defined on both nodes and links, if $\bm \sigma$ is defined only on links $\mathbf s=\mathbf s_{[1]}+\mathbf s_{[2]} $ will be defined on  nodes, links and triangles for simplicial complexes of dimension two.

\subsubsection{Noise Model}
We model the noise vector $\bm \epsilon$ as a standard Gaussian random vector, with elements drawn independently and identically at random:
\bea
{\mathbf x}\sim \mathcal{N}(0,{\bf I}).
\label{noise_gaussian1}
\eea
The noise vector $\bm\epsilon_{[n]}$ within each subspace $\mbox{im}(\mathbf D_{[n]})$ ($n=1,2$) can then be computed as:
\bea
{\bm \epsilon_{[n]}}=\alpha_n \frac{{\bf D}_{[n]}{\mathbf D}_{[n]}^{\dag} \mathbf x}{\sqrt{D_n}}
\label{noise_gaussian2}
\eea
where $D_n = |\mbox{im}(\mathbf D_{[n]})|$ is the dimension of the non-harmonic subspace of $\mathbf D_n$. 
This ensures that on expectation the noise is normalised to $\alpha_n$ on $\mbox{im}(\mathbf D_1)$ (i.e., nodes and links), or $\mbox{im}(\mathbf D_2)$ (i.e. links and triangles), respectively.
To quantify the noise within each subspace of the measured signal $\measured _n = \mathbf s_{[n]} +\bm \epsilon_{[n]}$, we define the signal to noise ratio as 
\begin{equation}
    \mbox{snr} = \frac{\|\mathbf s_{[n]}\|^2}{\|\bm\epsilon_{[n]}\|^2}
\end{equation}
Since $\|\mathbf s_{[n]}\|^2 = 1$ by construction, this quantity is in expectation $1/\alpha_n^2$.

\begin{figure*}[htb!]
\centering
\includegraphics[width=1\columnwidth]{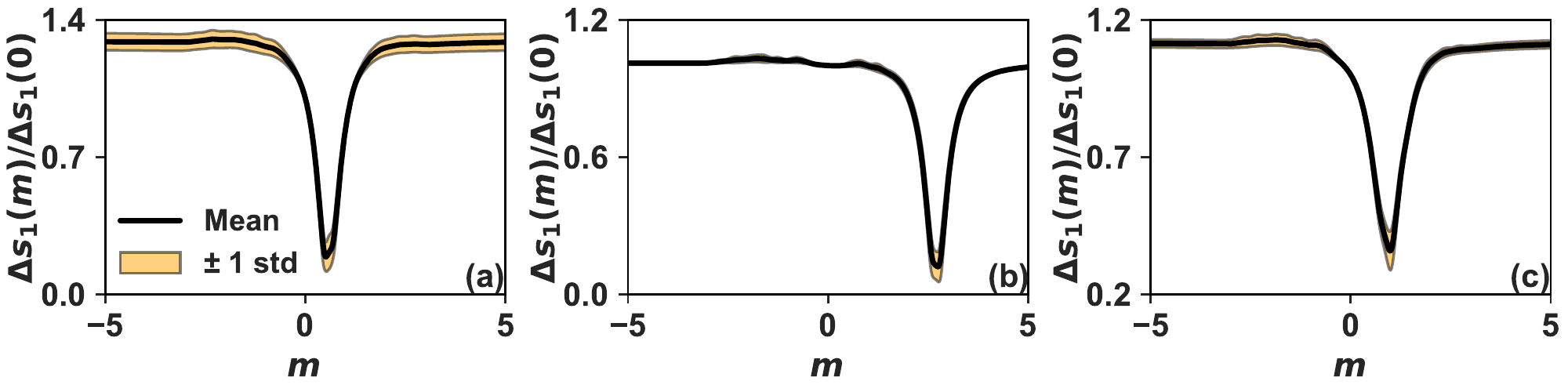}
\caption{The ratio between the error in the filtered signal, $\Delta s_{1}(m)$, and the error obtained with the Hodge Laplacian filtering algorithm $\Delta s_{1}(0)$, is shown  for different values of the  parameter $m$ on the Florentine Families network of marriage relations~\cite{Manlio_repository}. The true signal is given by the eigenvector with smallest (panel (a)) and largest (panel (b)) positive eigenvalue of the Dirac operator. Panel (c) presents the relative error in the case of true signal built as a linear combination  of eigenmodes with Gaussian coefficients centred at $\bar{\lambda}=1$ with standard deviation $\hat{\sigma} =0.2$. The ratio $\Delta s_{1}(m)/\Delta s_{1}(0)$ displays a minimum  where the parameter $m$ is equal to the true $m$ of the signal. The parameters are $\alpha_1=0.6$ and $\tau=10$. The relative error shown is averaged over 500 iterations, and the shaded region corresponds to the $\pm 1$ standard deviation interval around the mean.}
\label{fig:mass_sweep}
\end{figure*}
\begin{figure*}[!htb!]
\centering
\includegraphics[width=1\columnwidth]{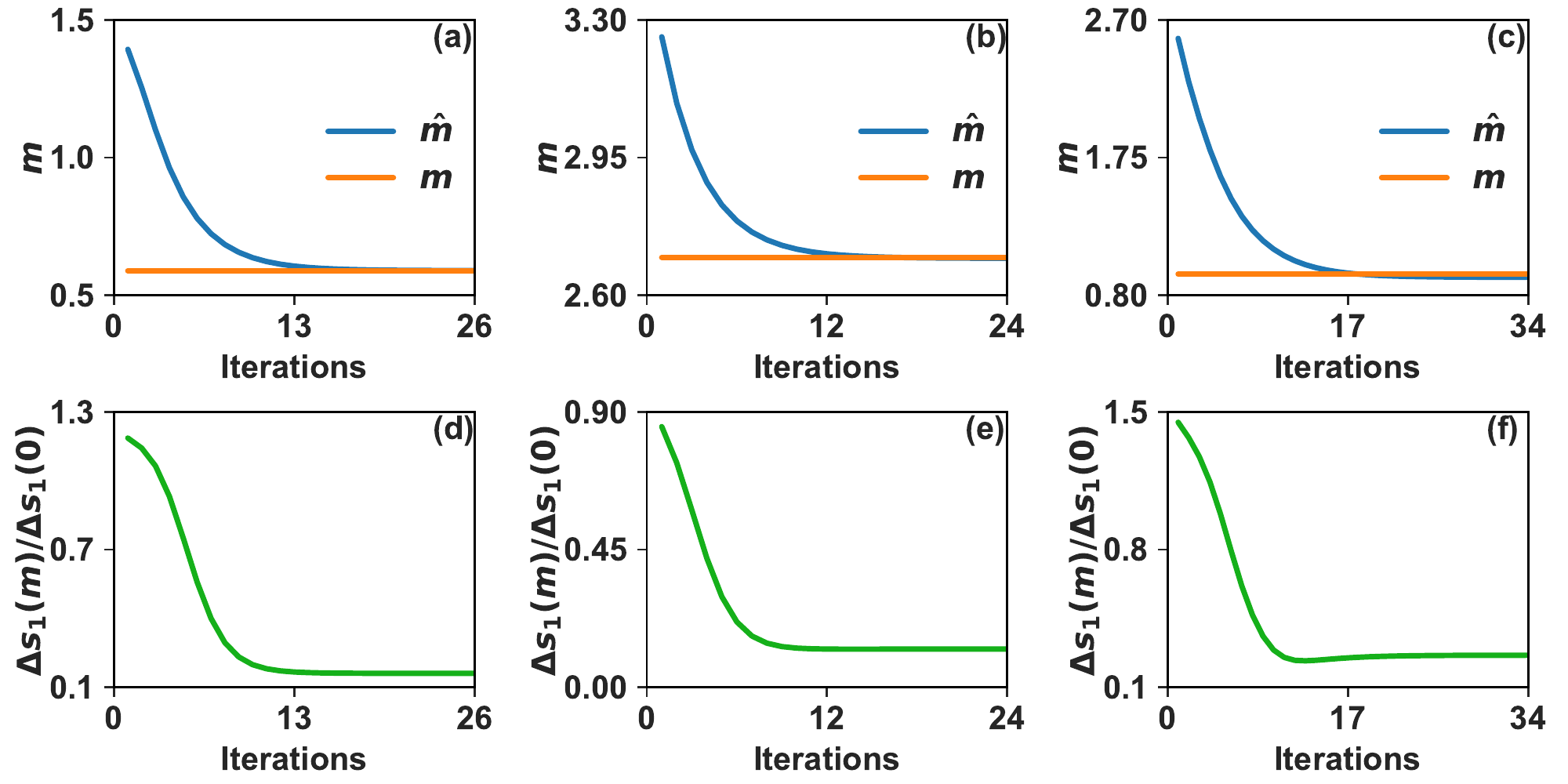}
\caption{We illustrate the learning process of the  {Dirac signal processing}  over consecutive iterations of the algorithm applied to  the Florentine Families network of marriage relations. Panels (a)-(c) compare the learned value $\hat{m}$ to the true $m$ of the pure signal during the learning progresses until the algorithm converges. Panels (d)-(f) present the relative error $\Delta s(m)/\Delta s (0)$, where $\Delta s (0)$ represents the error in the estimated signal with the Hodge Laplacian ($m=0$). This error greatly reduces over the optimization.
In panels (a) and (d), the true signal is given by the eigenvector of the Dirac operator with smallest (positive) eigenvalue in magnitude. In panels (b) and (e), it is given by the eigenvector with largest (positive) eigenvalue. In panels (c) and (f) the true signal is instead a linear combination of Dirac eigenmodes with Gaussian coefficients  centred at $\bar{\lambda} = 1$, with standard deviation $\hat{\sigma}=0.2$ of the Dirac eigenvectors. The parameters are $\alpha_1=0.5$, $\tau=7$ in panels (a), (b), (d), (e) and $\tau=2$ in panels (c) and (f), $\eta = 0.3$, $\delta = 10^{-4}$ and $m_1^{(0)} = 1.5$ in panels (a) and (d) and $m_1^{(0)} = 3$ in panels (b), (c), (e) and (d). The signal to noise ratio are respectively $5.15$, $4.52$ and $3.61$ in the first, second and third column.
}
\label{fig:mass_learning}
\end{figure*}

\section{Results}\label{sec:num_experiments}

\subsection{Application to the Florentine-Families network dataset}

To start with, we validate our algorithm on a network dataset,   i.e. a simplicial complex of dimension one, only formed by nodes and links. In particular, we consider the marriage layer of the Florentine Families multiplex network~\cite{Manlio_repository}.
This network is formed by $N_{[0]}=15$ nodes and $N_{[1]}=20$ links. The number of triads ($2$-simplices) is taken to be zero,   i.e. $N_{[2]}=0$ so that ${\bf D}_{[2]}={\bf 0}$.
We consider as true signal either the eigenvector corresponding to the largest positive  or  to the  smallest positive eigenvalue,
\bea
\mathbf s=\mathbf s_{[1]}= \bm \phi_1^{+}.
\eea

\begin{figure*}[!htb!]
\centering
\includegraphics[width=1\columnwidth]{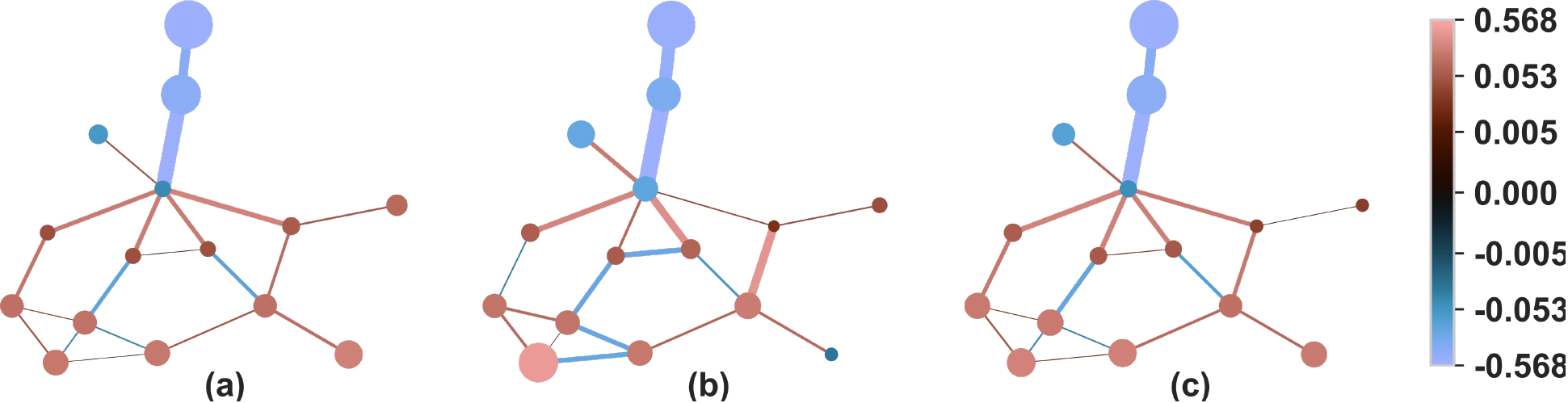}
\caption{Visualization of the true signal (panel (a)) the noisy signal (panel (b)) and the filtered signal (panel (c)) defined on nodes and links of the Florentine Family network~\cite{Manlio_repository}.  The true signal is given by the eigenvector of the Dirac operator with smallest eigenvalue. The parameter $\alpha_1$ is taken to be $0.6$, and the signal to noise ratio is $3.29$.
The filtered signal is obtained with $\tau=10$, $m_1 = 0.5$, $\eta = 0.1$ and 
$\delta = 10^{-4}$. The  error is $\Delta s_{1} = 0.14$. The nodes' sizes and links' widths are proportional to the local value of the signal considered. The magnitude  of the colorscale is logarithmic, except in the interval $[-0.005, 0.005]$ where it is linear.
}
\label{fig:visual_FF}
\end{figure*}
\begin{figure*}[!htb]
\centering
\includegraphics[width=1\columnwidth]{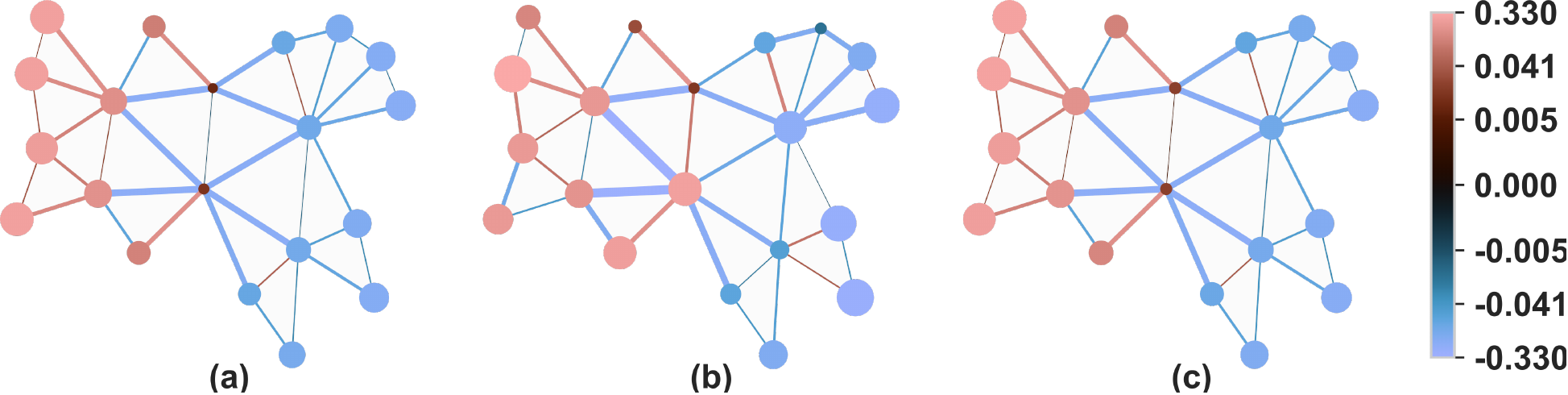}
\caption{Visualization of the true signal (panel (a)) the noisy signal (panel (b)) and the filtered signal (panel (c)) defined on the nodes and links of a single realization of the NGF network~\cite{gin_repository} with parameters $s_{NGF}=0$ and $\beta_{NGF}=0$. The true signal is given by the eigenvector of the Dirac operator with smallest eigenvalue. The parameter $\alpha_1$ is taken to be $0.7$, and the signal to noise ratio is $2.62$.
The filtered signal is obtained with $\tau=10$, $m_1 =1.1$, $\eta = 0.1$ and 
$\delta = 10^{-4}$. The  error obtained is $\Delta s_{1} = 0.06$. The nodes' sizes and links' widths are proportional to the local value of the signal considered. The magnitude of the  colorscale is logarithmic, except in the interval $[-0.005, 0.005]$ where it is linear.
}
\label{fig:visual_NL_NGF}
\end{figure*}
\begin{figure*}[!htb]
\centering
\includegraphics[width=1\columnwidth]{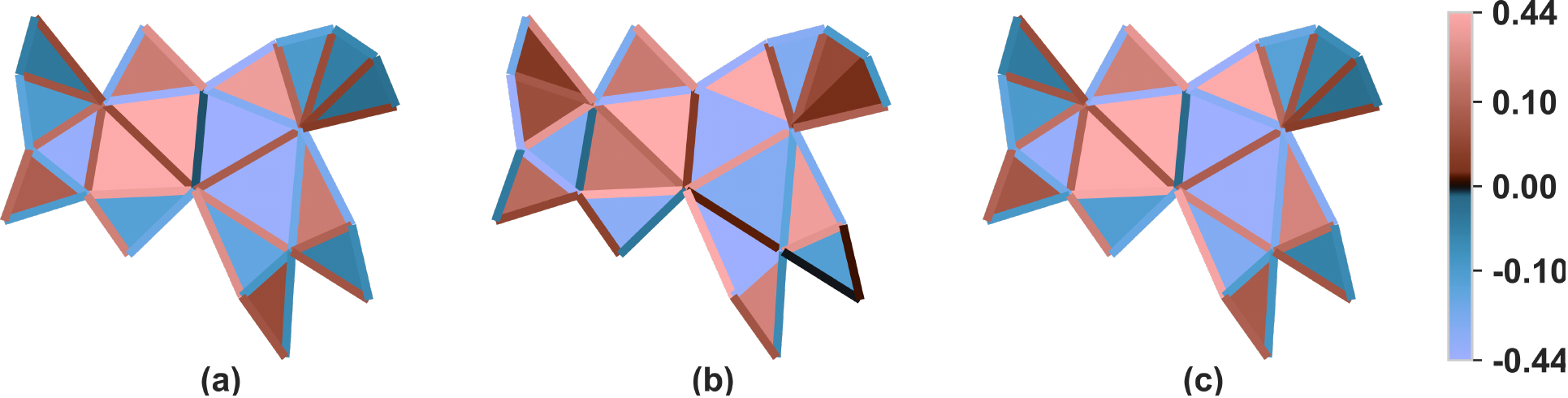}
\caption{Visualization of the true signal (panel (a)) the noisy signal (panel (b)) and the filtered signal (panel (c)) defined on the links  and triangles of a single realization of the NGF network~\cite{gin_repository} with parameters $s_{NGF}=0$ and $\beta_{NGF}=0$. The true signal is given by the eigenvector of the Dirac operator with smallest eigenvalue. The parameter $\alpha_2$ is taken to be $0.7$, and the signal to noise ratio is $2.16$.
The filtered signal is obtained with $\tau=10$, $m_2 = 1.6$, $\eta = 0.1$ and 
$\delta = 10^{-4}$. The  error obtained is $\Delta s_{2} = 0.1$. The magnitude of the colorscale is logarithmic, except in the interval $[-0.2, 0.2]$ where it is linear.}
\label{fig:visual_LT_NGF}
\end{figure*}
\begin{figure*}[!htb]
\centering
\includegraphics[width=1\columnwidth]{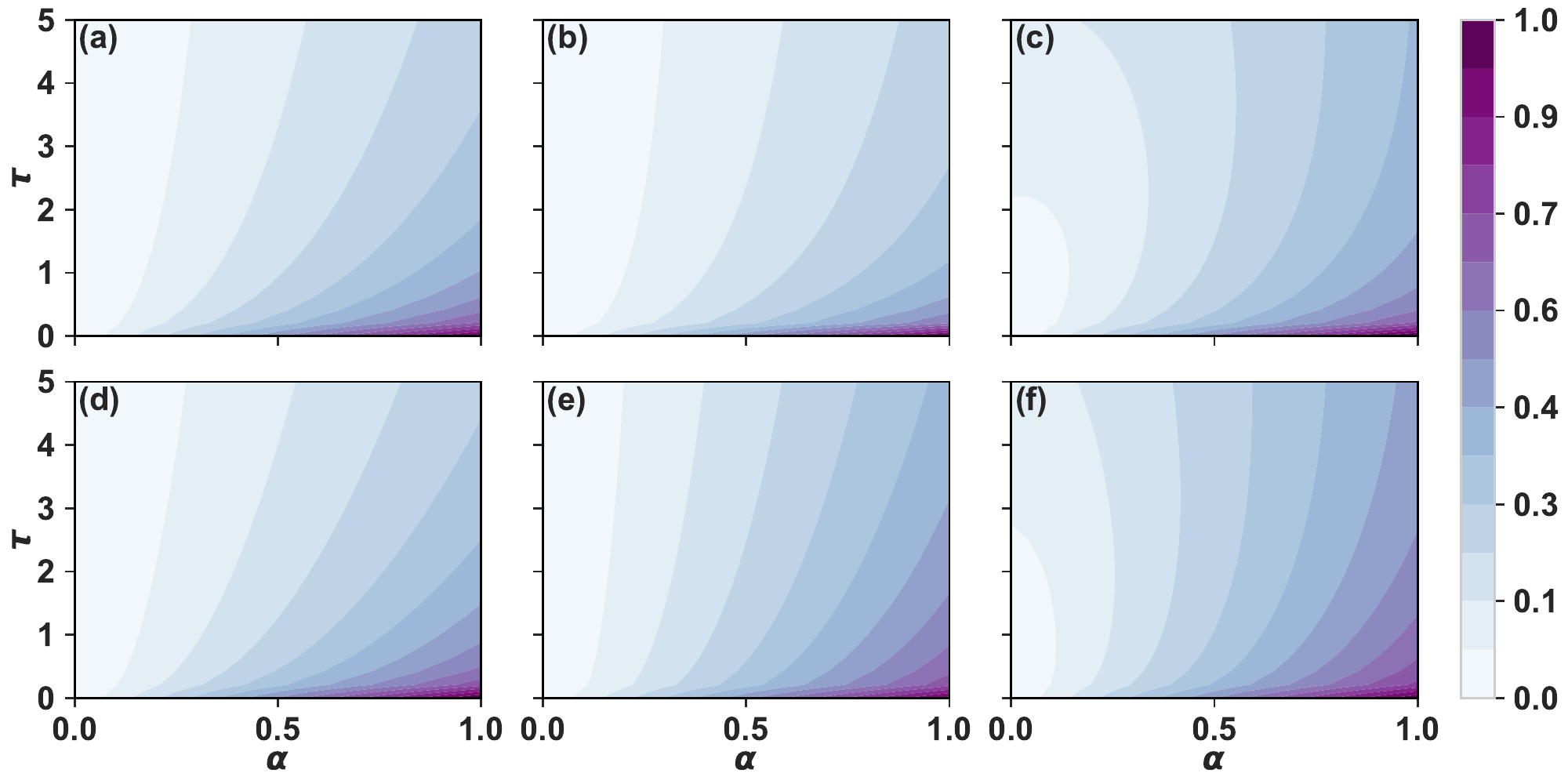}
\caption{The absolute error $\Delta s_{n}(m)$ of the reconstructed signal with optimised $m$ is shown as a function of $\tau$ and $\alpha_n$. Different panels correspond to the results obtained adopting different true signals. In panels (a) and (d) the true signal  is given by the eigenvector with (positive) smallest eigenvalue of respectively $D_{[1]}$ and $D_{[2]}$. In panels (b), (e), the true signal  is given by the eigenvector with (positive) largest eigenvalue of respectively $D_{[1]}$ and $D_{[2]}$. In panels (c) and (f) the true signal  is given by a linear combination  (with Gaussian coefficients) of eigenmodes of $D_{[1]}$ and $D_{[2]}$ eigenvectors respectively. In both cases the Gaussian coefficients are centred at $\bar{\lambda}=1$ and have standard deviation $\hat{\sigma}=0.2$. The error shown is averaged over $10$ realisations of noise. The parameters taken are $m_1^{(0)} = m_2^{(0)} =1$ in panels (a), (d), $m_1^{(0)} = m_2^{(0)} =3$ in panels (b), (e) and $m_1^{(0)} = m_2^{(0)} =2$ in panels (c), (f). The other parameters are $\eta = 0.3$ and $\delta = 10^{-4}$ for all panels.
}
\label{fig:NGF}
\end{figure*}

To illustrate the effect of the parameter $m_n$, we initially filter a noisy signal on this network by tuning the parameter $m_1=m$ rather than learning it, as shown in Fig. \ref{fig:mass_sweep}. 
The error $\Delta s_{1}(m)$  in the reconstructed signal defined in Eq. (\ref{eq:er}) shows a drastic dip for $m=\lambda$ where $\lambda$ is the eigenvalue corresponding to the true signal. 
Note that the minimal value of the error $\Delta s_{1}(m)$  obtained with the optimal choice of the parameter $m=\lambda\neq 0$ is much smaller than the error $\Delta s_{1}(m=0)$ that can be obtained with a standard signal processing algorithm using only the Hodge Laplacian   i.e. the algorithm obtained for $m=0$. 
This effect can be observed independently of the choice of the eigenvector chosen for the true signal (see Figure \ref{fig:mass_sweep}). 
Also considering a true signal formed by a linear combination of positive eigenvectors with Gaussian coefficients, can still lead to a well defined minimum of the relative error $\Delta s_{1}(m)/\Delta s_{1}(0)$ (see Figure $\ref{fig:mass_sweep}$c).

Furthermore, using the  {Dirac signal processing} algorithm, we can learn the value of parameter $m$. This is illustrated in Fig. \ref{fig:mass_learning} where we show the learned parameter $\hat{m}$ versus the true parameter $m$ as a function of the number of iterations of the algorithm,  for signals aligned with single positive eigenvectors, as well as constructed as linear combination  of different eigenvectors with Gaussian coefficients. In all cases, the proposed filtering is able to reconstruct the true signal much more closely than the Hodge-Laplacian filter, as revealed in the bottom row of Fig. \ref{fig:mass_learning}. Furthermore, the filtering proposed reconstructs the considered signals with significant error reduction compared to the noisy signal input. We find for the signal built as a linear combination of Dirac eigenmodes with Gaussian coefficients considered in Fig. \ref{fig:mass_learning}, the reduction reached is approximately $65\%$. For single eigenvectors, this furthermore can reach up to $90\%$.

This performance is illustrated visually in Fig. \ref{fig:visual_FF} where we compare in panel (a), (b) and (c) the true signal $\mathbf s_{[1]}$, the measured signal with random noise added and the reconstructed signal $\estimated_{[1]}$. The reconstructed signal shows excellent visual agreement with the true signal.
\begin{figure*}[!htb]
\centering
\includegraphics[width=1\columnwidth]{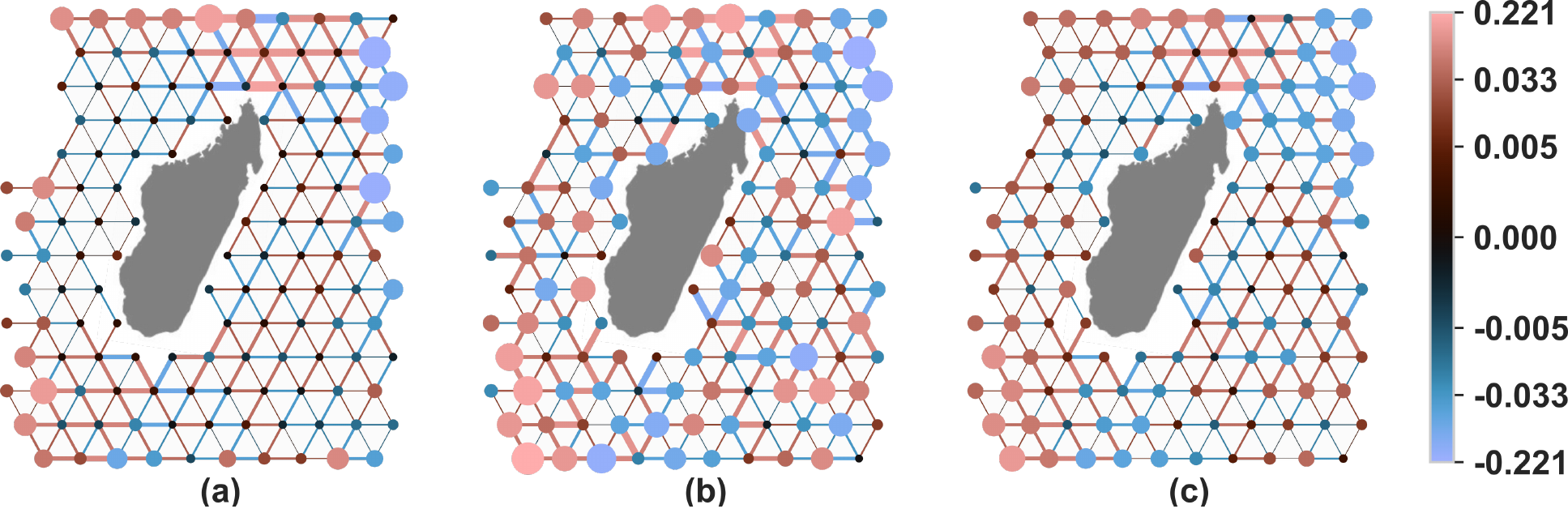}
\caption{Visualization of the true signal (panel (a)) the noisy signal (panel (b)) and the filtered signal (panel (c)) defined on the nodes  and links obtained from the drifter dataset on the shore of  Madagascar~\cite{drifter}. The parameter $\alpha_1$ is taken to be $1$, and the signal to noise ratio is $1.1$.
The filtered signal is obtained with $\tau=0.5$, $m_1^{(0)} = 1.1$, $\eta = 0.3$ and 
$\delta = 10^{-4}$. The obtained error is $\Delta s_{1} = 0.50$. The nodes' sizes and links' widths are proportional to the local value of the signal considered. The magnitude of the colorscale is logarithmic, except in the interval $[-0.005, 0.005]$ where it is linear. The island of Madagascar is shown for visualization purposes only.}
\label{fig:visual_NL_buoys}
\end{figure*}
\begin{figure*}[!htb]
\centering
\includegraphics[width=1\columnwidth]{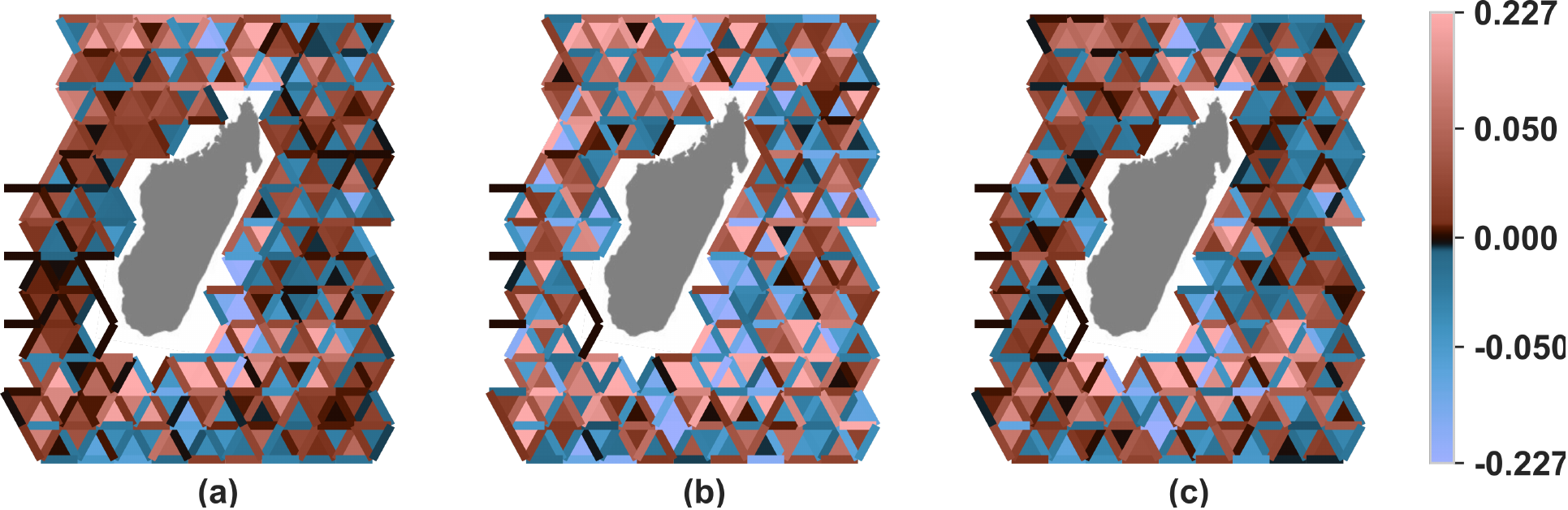}
\caption{ Visualization of the true signal (panel (a)) the noisy signal (panel (b)) and the filtered signal (panel (c)) defined on the  links and triangles obtained from the drifter dataset on the shore of  Madagascar~\cite{drifter}.  The parameter $\alpha_2$ is taken to be $1$, and the signal to noise ratio is $1.03$.
The filtered signal is obtained with $\tau=1$, $m_2^{(0)} = 1.6$, $\eta = 0.3$ and 
$\delta = 10^{-4}$. The obtained error is $\Delta s_{2} = 0.64$. The magnitude of the colorscale is logarithmic, except in the interval $[-0.1, 0.1]$ where it is linear. The island of Madagascar is shown for visualization purposes only.}
\label{fig:visual_LT_buoys}
\end{figure*}
\subsection{Application to the simplicial complex model NGF}

The Network Geometry with Flavor (NGF) \cite{bianconi2016network,bianconi2017emergent,bianconi2021higher} is a very comprehensive model of growing simplicial complexes able to generate discrete manifolds and more general structures, whose $1$-skeleton, (i.e. the network obtained by retaining only the nodes and links of the simplicial complex) has high clustering coefficient and high modularity. 
In certain limits the NGF reduces to existing models such as the Barabasi-Albert model or the Apollonian random graph model.
Here we consider an NGF dataset of dimension $d=2$   i.e. formed by nodes, links and triangles which is a discrete manifold and can be generated by setting the NGF model parameters to $s_{NGF}=-1$ (flavor) and $\beta_{NGF}=0$ (inverse temperature). 
This generates discrete manifolds of dimension $d=2$ with exponential degree distributions. 
The code to generate the NGF simplicial complexes of any arbitrary dimension is freely available in the repository~\cite{gin_repository}.

The simplicial complex being two dimensional, we can consider a true signal defined on nodes, links and triangles and test our complete algorithmic pipeline.
In particular we take the true signals $\mathbf s_{[1]}$ and $\mathbf s_{[2]}$ given by Eq. \ref{eq:signals_eig} with $c_1=c_2=1$ and proportional to an arbitrary eigenvector associated  to $\mathbf D_{[1]}$ and $\mathbf D_{[2]}$ respectively. 
In  Figs. \ref{fig:visual_NL_NGF} and \ref{fig:visual_LT_NGF} we visualise  the performance of our smoothing algorithm when the parameter $m$ is learned for signals supported by nodes and links, and links and triangles respectively. In both cases, it is visually clear that our filtering approach can reconstruct the signal considered (in panel (c)) very closely to the true signal (in panel (a)) from the noisy measurement shown in panel (b).

In Figure $\ref{fig:NGF}$, we report the performance of the  {Dirac signal processing} algorithm by measuring the error $\Delta s_{n}(\tau, \alpha)$ for $n\in \{1,2\}$ defined as  
{\bea
\Delta s_{n}=\|\estimated_{[n]}(\tau, \alpha) - \mathbf s_{[n]}\|_2,
\eea}
as a function of {the parameters} $\tau$ and $\alpha_n=\alpha$.
We measure this error over $10$ different iterations of the noise, and report the mean in Fig. \ref{fig:NGF}. For all single eigenvectors considered as well as signals consisting of a linear combination of of eigenvectors with Gaussian coefficients, we find that the algorithm yields errors around $0.1-0.3$ for all values of $\alpha$ considered provided $\tau$ is large enough.
{Finally we also observe that the  NGF model provides also a very solid benchmark to test the complexity of our algorithm (see for details Appendix B).}

\subsection{Application to the drifter dataset}
We test the  {Dirac signal processing} algorithm on the real dataset of drifters in the ocean from the
Global Ocean Drifter Program available at the AOML/NOAA Drifter Data Assembly Center~\cite{drifter}. The drifters data set already analyzed in Ref.~\cite{schaub2020random} consists of the individual trajectories of $339$ buoys around the island of Madagascar in the pacific ocean. Projected onto a tessellation of the space, this yields $339$ edge-flows{, each representing the motion of a buoy between pairs of cells.} The underlying simplicial complex itself consists of $133$ nodes, $322$ links and $186$ triangles. {In order to obtain a single topological signal spanning all links in the simplicial complex, we consider the link-wise sum of the $339$ trajectories, which represent the net physical flow around the island. This signal on links can further be encoded in} $\bm \sigma${, a} topological spinor {with zero values on nodes and triangles} of the $2$-dimensional simplex defined by the geographical tessellation. {A signal $\mathbf s$ spanning all three dimensions using Eq. (\ref{eq:dr}). Physically, the generated signals on nodes and triangles then each respectively encode node wise source/sinks of the flow (discrete divergence), and local circulation around $2-$dimensional triangles (discrete curl)}. {Finally, u}sing Eq. \ref{projection}, we obtain our signals $\mathbf s_{[1]}$ and $\mathbf s_{[2]}$, which we normalise to have unit norm. Noise vectors $\bm \epsilon_{[1]}$ and $\bm \epsilon_{[2]}$ obey Eq.(\ref{noise_gaussian2}).
\begin{figure}[!htb]
\centering
\includegraphics[width=0.7\columnwidth]{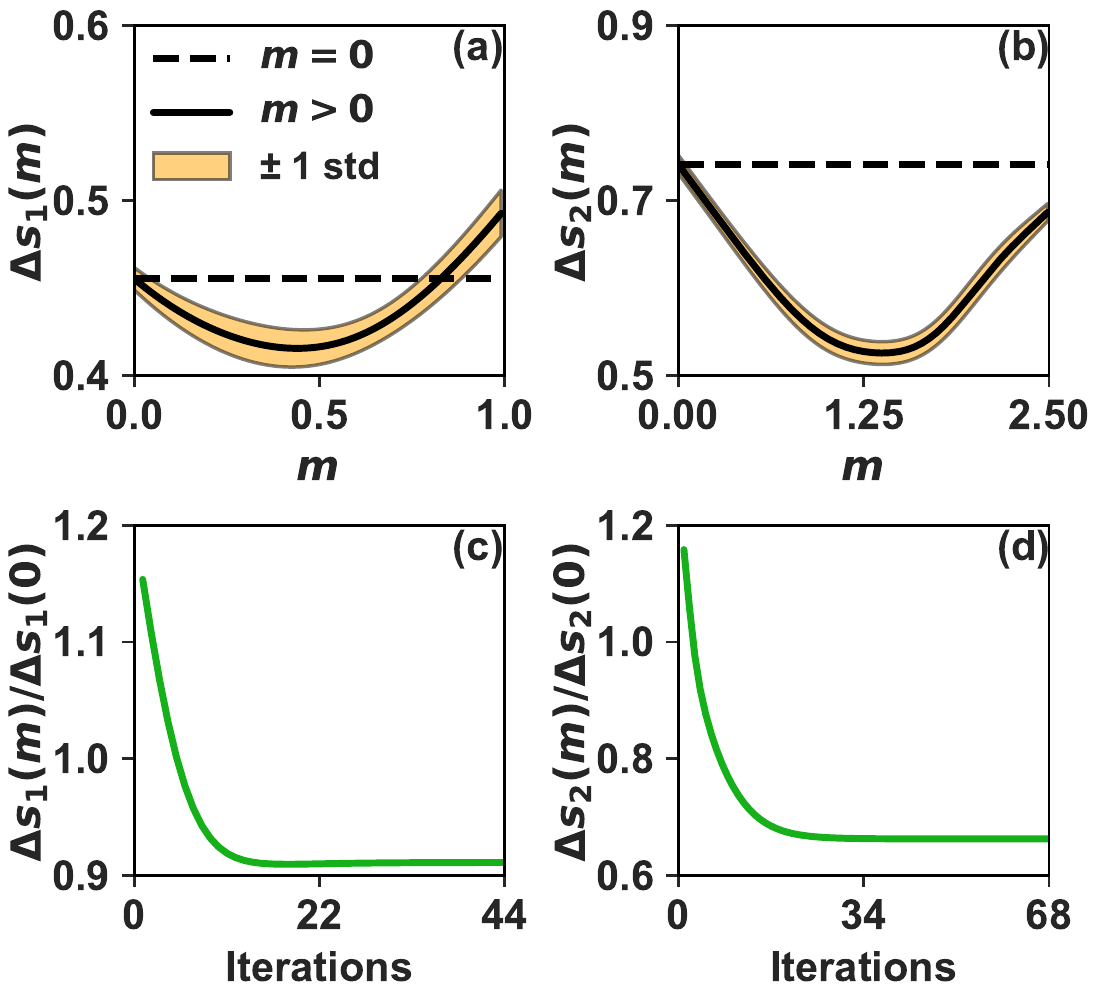}
\caption{The relative error in the filtered signal  of the data of drifters of the ocean \cite{drifter,schaub2020random} is shown over different values of $m$, in comparison to the error obtained with a Hodge-Laplacian filtering ($m=0$). The curve shown is averaged over $10$ realisations of noise. The region around the mean bounded by $1$ standard deviation is shaded. The algorithm is applied to the nodes-links signal in panel (a) and to links-triangles signal in panel (b). In panels (c) and (d) we learn the parameter $m$ and show the relative error over the learning process for a single iteration. The  {Dirac signal processing} algorithm whose performance is shown in panels (c) and (d) is able to reduce significantly the error with respect to the Hodge-Laplacian filtering, however we do not observe convergence to the true value of $m$. The parameters are $\alpha_n=0.6$, $\tau=1$ in panels (a) and (c); $\tau = 1.5$ in panels (b) and (d) and  $\eta = 0.3$, 
$\delta = 10^{-4}$ in panels (c) and (d).}
\label{fig:mass_sweep_buoys}
\end{figure}
 In Fig. \ref{fig:visual_NL_buoys}, we show in panel (a), (b) and (c) respectively the signal $\mathbf s_{[1]}$ we aim to reconstruct, the noisy signal $\measured_{[1]}$ and the filtered signal $\estimated_{[1]}$ with optimized $m$. We see visually  that the reconstructed signal is closer to the true signal than the measured noisy signal, confirming that  {Dirac signal processing} is able to filter key features of the signal, and leave out noisy components. This can be similarly observed on the signal projected onto links and triangles as shown in Fig. \ref{fig:visual_LT_buoys}.

The performance of the algorithm on this real data set can be evaluated by measuring the error for given value of  $m$. This is reported in panels (a) and (b) of Fig. \ref{fig:mass_sweep_buoys} together with the error obtained with a Hodge-Laplacian filtering. We observe in both cases an improvement of the  {Dirac signal processing} over the Hodge Laplacian signal processing corresponding to $m=0$, over   a significant range or values of $m$. This improvement of the  {Dirac signal processing} algorithm is more pronounced for the signals supported on links and triangles. 

We moreover find that the unsupervised version of the  {Dirac signal processing} algorithm where $m$ is a learned parameter is efficient in this real scenario (see Fig. \ref{fig:mass_sweep_buoys}). 
Indeed for  both $\mathbf s_{[1]}$ and $\mathbf s_{[2]}$ considered, we see that the ratio $\Delta s_{n}({m})/\Delta s_{n}(0)$ converges below $1$, showing an improvement in reconstruction when using the Dirac operator compared to the algorithm using exclusively the Hodge Laplacian.

On average, this relative error reaches $0.85$ and $0.65$ respectively for nodes and links, and links and triangles signals.

\section{Conclusions}\label{sec:conclusions}
In this paper we have proposed the  use of the Dirac operator  to jointly process topological signals on simplicial complexes, and demonstrated the utility of theses ideas by presenting the  {Dirac signal processing} algorithm to treat signals supported on nodes, links and triangles of simplicial complexes. 
Our algorithm exploits the chiral symmetry of the Dirac operator,  i.e. the fact that for each positive eigenvalue there is a corresponding negative eigenvalue whose eigenvectors are chiral.
Our proposed algorithm is furthermore adaptive in that it is able to learn the value of its parameter $m$ to efficiently filter the higher-order signals.
We have tested this algorithm on both synthetic data of networks and simplicial complexes and on real data of drifting buoys in the ocean. 
We found good performance of the algorithm under very general conditions of the signals.
Furthermore, we have demonstrated a significant improvement with respect to a corresponding algorithm using simply the Hodge Laplacian, showing that the Dirac operator can be a key ingredient to improve the filtering of topological data.
We believe that this work opens new perspectives for the use of the Dirac operator for machine learning of topological signals and hope that it can inspire further work between topology and machine learning along this or other relevant research directions. 
For instance, an interesting direction for further exploration is the use of directional Dirac operators to distinguish between links of different types of directions in a given network~\cite{dirac}.

 
\section*{Data availability.}
The Florentine Families multiplex network is available from \cite{manlio}.
The drifter data are extracted from Global Ocean Drifter Program available at the AOML/NOAA Drifter Data Assembly Center~\cite{drifter}.

\section*{Code availability.}
The code to generate the simplicial complex ``Network Geometry with Flavor" \cite{bianconi2016network} is freely available at the repository \cite{gin_git}. All other codes used in this work are available upon request.
\section*{Appendix A}

\begin{figure}[!htb]
\begin{center}
\includegraphics[width=0.6\columnwidth]{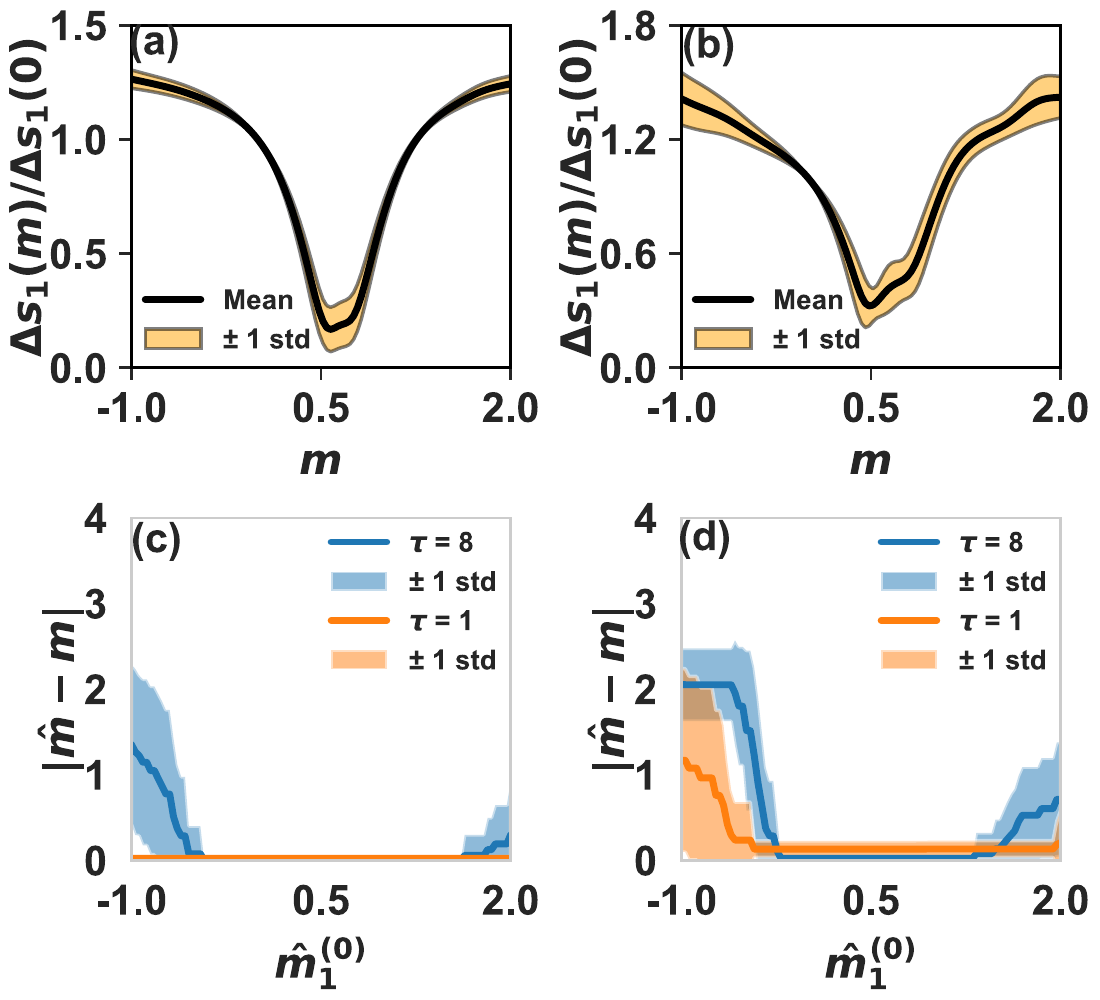}
\end{center}
\caption{{The average error in filtering the signal is shown relative to the Hodge Laplacian filter for a various range of fixed $m$ in panels (a) and (b) on the node and links of a simplicial complex generated from the NGF model. Instead in panels (c) and (d), the parameter $m$ is iteratively learned (denoted by $\hat{m}$), and the average difference with the true $m$ is instead plotted as a function of the initial guess used, denoted $\hat{m}_1^{(0)}$ for two values of $\tau$. In panels (a) and (c), the parameter $\alpha_1$ is set to $0.6$, and the iterative process converges well for most $\hat{m}_1^{(0)}$. In panels (b) and (d), $\alpha_1 = 1.5$ instead and the region of convergence is reduced. For all panels, the true signal is given by the eigenvector of $D_{[1]}$ with positive minimum eigenvalue, and $20$ realisations of noise are considered. In panels (c) and (d), $\eta = 0.3$ and $\delta = 10^{-4}$.}}
\label{fig:NGF_conv}
\end{figure}

{In this Appendix we investigate the convergence of our algorithm when the signal to noise ratio is reduced. To this end, we have compared the performance of the algorithm for two signal processing problems with a different level of the signal to noise ratio (see Fig. \ref{fig:NGF_conv}). 
In order to describe the difficulty of the task, we have plotted the error landscape of the algorithm obtained by measuring the normalized error of the algorithm when  the value of  $m$ is externally tuned (see panels (a) and (b) of Fig.\ref{fig:NGF_conv}). We observe clearly that the error landscape is more rough for lower signal to noise ratio implying that the reconstruction problem becomes more hard.
In this setting, we measured the absolute difference between the true value of $m$ and the  learned value, $\hat{m}$, as a function of the  initial guess, indicated by $\hat{m}_1^{(0)}$ (see panels (c) and (d) of Fig.\ref{fig:NGF_conv}). 
We observe that the convergence to the true value of $m$ is impacted by the choice of the initial condition $\hat{m}_1^{(0)}$ in addition to the signal to noise ratio. In particular, for lower signal to noise ratio an initial condition $\hat{m}_1^{(0)}$ closer to the true value is required to have good convergence properties.}

{In either cases, as a rule of thumb, we find that lowering the value of the parameter $\tau$   i.e. reducing the steepness of the filter (see discussion above) can alleviate the severity of this issue. Furthermore, introducing stochasticity within the algorithm in order to escape such local {extrema} may improve the filtering performance. However this generalization of the algorithm is beyond the scope of this work and might be the subject of some future extension of Dirac signal processing. We note however that the algorithm preserves good convergence for a wide range of parameters as shown in Fig. \ref{fig:NGF}.}

\section*{Appendix B}
\begin{figure}[!htb]
\begin{center}
\includegraphics[width=0.6\columnwidth]{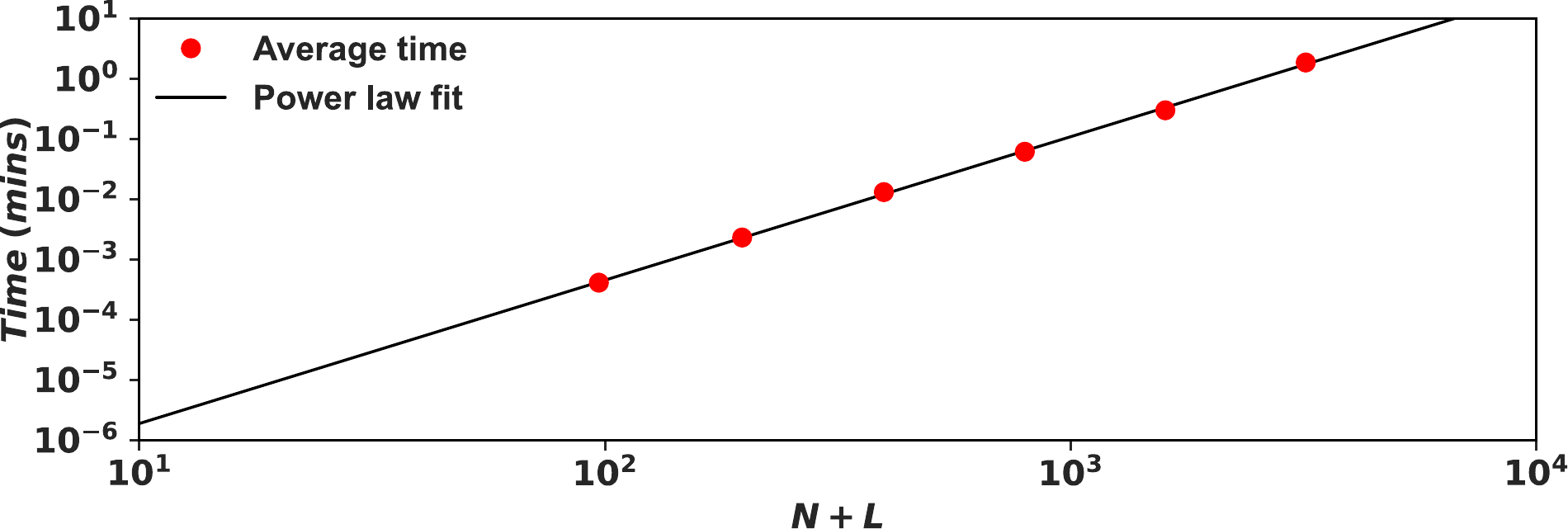}
\end{center}
\caption{{The average optimization time (algorithm runtime) is shown for signals on the node and links of simplicial complexes generated from the NGF model with increasing sizes. For each simplicial complex size, the time measured is averaged over the optimization of $20$ pure and noisy signals. Each signal considered is given by a linear combination (with Gaussian coefficients centred at $\bar{\lambda}=1$ with standard deviation $\hat{\sigma}=0.2$) of eigenvectors of $D_{[1]}$. The initial parameter $m_1^{(0)}$ is here given by a Rayleigh coefficient using the observed data. The other optimization parameters are $\eta = 0.3$ and $\delta = 10^{-4}$. The data can be fitted to a power law with scaling $2.38 \pm 0.026$.}}
\label{fig:NGF_perf}
\end{figure}

{In this section we investigate the complexity of the proposed Dirac signal processing algorithm, as implemented in our (not optimized) code.
In particular in Fig. \ref{fig:NGF_perf} we show the average time taken for the algorithm to converge as a function of the size of the NGF simplicial complex. We focus in particular on the complexity of the algorithm filtering  signals in the image of $D_{[1]}$ localized on nodes and links. This approximately follows a power law scaling in $N+L$, and a simple fit to such a model evaluates the scaling exponent to $2.38 \pm 0.026$. For this analysis, signals on simplicial complexes of up to $1600$ nodes (equivalent to signals of dimension $N+L = 3197$) were processed, each taking under $2$ minutes. This confirms that in its current implementation, the algorithm is applicable to data sets of practically relevant sizes. Furthermore, the measured time was obtained on a local computer, and we expect further work focused on optimising the method to improve this performance.}
\section*{References}
\bibliographystyle{unsrt}
\bibliography{references_paper}

\end{document}